\RequirePackage{fix-cm}
\documentclass[twocolumn,epjc3]{svjour3}
\smartqed  
\usepackage{amsmath}
\usepackage{amssymb}
\usepackage{amssymb,bbold}
\usepackage{kantlipsum,widetext}

\RequirePackage{graphicx}
\RequirePackage[numbers,sort&compress]{natbib}

\usepackage{lineno,hyperref}
\hypersetup{colorlinks=true,linkcolor=blue,citecolor=blue,filecolor=blue,urlcolor=blue,breaklinks=true}

\journalname{Eur. Phys. J. C}
\begin{document}

\title{Noninertial effects on the quantum dynamics of scalar bosons
}


\author{Luis B. Castro\thanksref{e1,addr1} 
}

\thankstext{e1}{e-mail: lrb.castro@ufma.br, luis.castro@pq.cnpq.br}


\institute{Departamento de F\'{\i}sica, Universidade Federal do Maranh\~{a}o, Campus Universit\'{a}rio do Bacanga, 65080-805, S\~{a}o Lu\'{\i}s, MA, Brazil.
\label{addr1}
}

\date{Received: date / Accepted: date}

\maketitle

\begin{abstract}
The noninertial effect of rotating frames on the quantum dynamics of scalar bosons embedded in the background of a cosmic string is considered. In this work, scalar bosons are described by the Duffin--Kemmer--Petiau (DKP) formalism. Considering the DKP oscillator in this background the combined effects of a rotating frames and cosmic string on the equation of motion, energy spectrum and DKP spinor are analyzed and discussed in details. Additionally, the effect of rotating frames on the scalar bosons localization is studied.

\end{abstract}

\section{Introduction}
\label{intro}

The Duffin-Kemmer-Petiau (DKP) formalism  \cite{Petiau1936,Kemmer1938,PR54:1114:1938,Kemmer1939} is a first-order relativistic equation that describes spin-zero and spin-one particles and has been used to analyze relativistic interactions of spin-zero and spin-one hadrons with nuclei as an alternative to their conventional second-order Klein-Gordon (KG) and Proca counterparts. Although the formalisms are equivalent in the case of minimally coupled vector interactions \cite{PLA244:329:1998,PLA268:165:2000,PRA90:022101:2014}, the DKP formalism enjoys a richness of couplings not capable of being expressed in the KG and Proca theories \cite{PRD15:1518:1977,JPA12:665:1979}. Recently, there has been an increasing interest on the so-called DKP oscillator \cite{ZPC56:421:1992,JPA27:4301:1994,JPA31:3867:1998,JPA31:6717:1998,PLA346:261:2005,
JMP47:062301:2006,PS76:669:2007,PS78:045010:2008,JMP49:022302:2008,
IJTP47:2249:2008}. The DKP oscillator considering minimal length \cite{JMP50:023508:2009,
JMP51:033516:2010}, noncommutative phase space \cite{CTP50:587:2008,CJP87:989:2009,IJTP49:644:2010,EPJC72:2217:2012} and topological defects \cite{EPJC75:287:2015} have also appeared in the literature. The DKP oscillator is a kind of tensor coupling with a linear potential which leads to the harmonic oscillator problem in the weak-coupling limit. Also, a sort of vector DKP oscillator (non-minimal vector coupling with a linear potential \cite{MPLA20:43:2005,JPA43:055306:2010,NPBPS199:203:2010,PLA375:2596:2011,
ADHEP2014:784072:2014}) has been an topic of recent investigation. Vector DKP oscillator is the name given to the system with a Lorentz vector coupling which exhibits an equally spaced energy spectrum in the weak-coupling limit.
The name distinguishes from that system called DKP oscillator with Lorentz tensor couplings of Ref.~\cite{ZPC56:421:1992,JPA27:4301:1994,JPA31:3867:1998,JPA31:6717:1998,PLA346:261:2005,
JMP47:062301:2006,PS76:669:2007,PS78:045010:2008,JMP49:022302:2008,
IJTP47:2249:2008,JMP50:023508:2009,JMP51:033516:2010,CTP50:587:2008,CJP87:989:2009,IJTP49:644:2010,
EPJC72:2217:2012}.

The DKP oscillator is an analogous to Dirac oscillator \cite{JPA22:L817:1989}. The Dirac oscillator is a natural model for studying properties of physical systems, it is an exactly solvable model, several research have been developed in the context of this theoretical framework in recent years. A detailed description for the Dirac oscillator is given in Ref.~\cite{STRANGE1998} and for other contributions see Refs.~\cite{PLA325:21:2004,JPA39:5125:2006,PRC73:054309:2006,
PRA84:032109:2011,AP336:489:2013,PLB731:327:2014,AP356:83:2015}. The Dirac oscillator embedded in a cosmic string background has inspired a great deal of research in last years\cite{NPB328:140:1989,PRL62:1071:1989,PLA361:13:2007,PRD83:125025:2011,
PRD85:041701:2012,AP339:510:2013,EPJC74:3187:2014}. A cosmic string is a linear defect that change the topology of the medium when viewed globally. The influence of this topological defect in the dynamics of spin-$1/2$ particles has been widely discussed in the literature. 

On the other hand, the standard description of physical phenomena according to accelerated observers is based on a hypothesis of locality which states that an accelerated observer at each instant along its wordline is equivalent to a hypothetical inertial observer at the same event and with the same velocity as the noninertial observer. This assumption forms the basis for the extension of the Poincar\'{e}--invariant theory of relativity to general frames of references as well as gravitational fields. The study of rotating frames has discovered interesting effects, where the best-known effect is the Sagnac effect \cite{CRAS157:708:1913,CRAS157:1410:1913}. Another known effect is the Mashhoon effect \cite{PRL61:2639:1988}, which yields a phase shift due to the coupling between the spin of the particle with the angular velocity of the rotating frames. Also we have the Page--Werner et al. term \cite{PRL35:543:1975,PRL42:1103:1979}, which is a coupling between the angular momentum of the particle and the angular velocity of the rotating frame. Other studies of noninertial effects in quantum systems have also been extended to confined systems, for instance rotational and gravitational effects in quantum interference \cite{PRD15:1448:1977,PRD20:1846:1979,PLA195:284:1994}, scalar fields \cite{PRD22:1345:1980,PRD85:061502:2012}, Dirac fields \cite{PRD26:1900:1982}, persistent currents in quantum rings \cite{PLA197:444:1995}, confinement of a neutral particle to a quantum dot \cite{PLA374:4642:2010,MPLB27:1350018:2013}, Dirac oscillator \cite{EPJP127:82:2012,GRG45:1847:2013},  and spin currents \cite{PLA377:960:2013}. Recently, the noninertial effects due to rotation or acceleration have been investigated in condensed matter systems as for example, noninertial effects due to a rotating hall sample \cite{EPL54:502:2001}, rotating Bose--Einstein (BE) condensation in ultra could diluted atomic gases \cite{PRA76:023410:2007}, effect of rotating frame in $C_{60}$ molecules \cite{PRB68:195421:2003,EPJD33:35:2005}, among others systems. However, investigations on noninertial effects involving scalar bosons via the DKP formalism has not been established,  therefore we believe that this problem deserves to be explored.

The main motivation of this work is to study the noninertial effects on the quantum dynamics of scalar bosons embedded in the background of a cosmic string. In this work, the influence of combined effects of the angular velocity of the rotating frame $\varpi$ and the angular deficit of the cosmic string $\alpha$ in the equation of motion, energy spectrum and DKP spinor are analyzed and discussed in details. The case of DKP oscillator in this background is also considered. Owing to peculiar behavior of this background, one can readily envisage that two different classes of solutions can be segregated depending on the value of the product $\varpi\alpha$. For an arbitrary value of $\varpi\alpha$ and considering the aproppriate boundary condition, the possible energy levels are obtained by a root-finding procedure of a symbolic algebra program. On the other hand, for the limit $\varpi\alpha\ll 1$ and considering the appropriate bounday condition, the exact solutions are presented in closed form.
We show that scalar bosons and antibosons tend to be better localized when the angular velocity of the rotating frame $\varpi$ increases. The results reported in \cite{EPJC75:287:2015} can be obtained as a particular case.

This work is organized as follows. In section \ref{sec:1}, we consider a short review on DKP equation in a curved space-time. In section \ref{sec:2}, we give a brief review on a cosmic string background, noninertial reference frame and we also analyze the curved-space beta matrices and spin connection in this background. In section \ref{sec:3}, we concentrate our efforts in the interaction called DKP oscillator embedded in the background of a cosmic string in a rotating coordinate system. In particular, we focus the case of scalar bosons and obtain the equation of motion, energy spectrum and DKP spinor. We analyze two kinds of solutions that depend on the value of the product $\varpi\alpha$. Finally, in section \ref{sec:4} we present our conclusions.

\section{Review on Duffin--Kemmer--Petiau equation in a curved space-time}
\label{sec:1}

The Duffin-Kemmer-Petiau (DKP) equation for a free boson in curved space-time is given by \cite{GRG34:491:2002,
GRG34:1941:2002} ($\hbar =c=1$)%
\begin{equation}\label{dkp}
\left[ i\beta ^{\mu }\nabla_{\mu}-M\right] \Psi =0
\end{equation}%
\noindent where the covariant derivative
\begin{equation}\label{der_cov}
\nabla_{\mu}=\partial _{\mu }-\Gamma_{\mu}\,.
\end{equation}
\noindent In this case, we restrict our analysis to the torsion--zero case and the affine connection is defined by
\begin{equation}\label{affine}
\Gamma_{\mu}=\frac{1}{2}\,\omega_{\mu\bar{a}\bar{b}}[\beta^{\bar{a}},\beta^{\bar{b}}]\,.
\end{equation}
\noindent The curved-space beta matrices are
\begin{equation}\label{beta_curved}
\beta ^{\mu }=e^{\mu}\,_{\bar{a}}\,\beta^{\bar{a}}
\end{equation}
\noindent and satisfy the algebra%
\begin{equation}\label{betaalge}
\beta ^{\mu }\beta ^{\nu }\beta ^{\lambda }+\beta ^{\lambda }\beta
^{\nu }\beta ^{\mu }=g^{\mu \nu }\beta ^{\lambda }+g^{\lambda \nu
}\beta ^{\mu }\,.
\end{equation}%
\noindent where $g^{\mu \nu }$ is the metric tensor. The algebra expressed by (\ref{betaalge}) generates a set of 126 independent matrices whose irreducible representations are a trivial representation, a five-dimensional representation describing the spin-zero particles (scalar sector) and a ten-dimensional representation associated to spin-one particles (vector sector). More detailed discussions on the DKP formalism in a curved space-time can be found in Ref. \cite{EPJC75:287:2015}.

The \textit{tetrads} $e_{\mu}\,^{\bar{a}}(x)$ satisfy the relations%
\begin{equation}\label{tetr1}
\eta^{\bar{a}\bar{b}}=e_{\mu}\,^{\bar{a}}\,e_{\nu}\,^{\bar{b}}\,g^{\mu\nu}
\end{equation}%
\begin{equation}\label{tetr2}
g_{\mu\nu}=e_{\mu}\,^{\bar{a}}\,e_{\nu}\,^{\bar{b}}\,\eta_{\bar{a}\bar{b}}
\end{equation}%
\noindent and
\begin{equation}\label{tetr3}
e_{\mu}\,^{\bar{a}}\,e^{\mu}\,_{\bar{b}}=\delta^{\bar{a}}_{\bar{b}}
\end{equation}%
\noindent the Latin indexes being raised and lowered by the Min\-kowski metric tensor $\eta^{\bar{a}\bar{b}}$ with signature $(-,+,+,+)$ and the Greek ones by the metric tensor $g^{\mu\nu}$.

\noindent The spin connection $\omega_{\mu\bar{a}\bar{b}}$ is given by
\begin{equation}\label{con}
\omega_{\mu}\,^{\bar{a}\bar{b}}=e_{\alpha}\,^{\bar{a}}\,e^{\nu \bar{b}}\,\Gamma_{\mu\nu}^{\alpha}
-e^{\nu \bar{b}}\partial_{\mu}e_{\nu}\,^{\bar{a}}
\end{equation}%
\noindent with $\omega_{\mu}\,^{\bar{a}\bar{b}}=-\omega_{\mu}\,^{\bar{b}\bar{a}}$ and $\Gamma_{\mu\nu}^{\alpha}$ are the Christoffel symbols given by
\begin{equation}\label{symc}
\Gamma_{\mu\nu}^{\alpha}=\frac{g^{\alpha\beta}}{2}\left( \partial_{\mu}g_{\beta\nu}+
\partial_{\nu}g_{\beta\mu}-\partial_{\beta}g_{\mu\nu} \right).
\end{equation}%

\noindent As it shown in Ref. \cite{EPJC75:287:2015}, the conservation law for $J^{\mu}$ leads to
\begin{equation}\label{corr}
\nabla_{\mu}J^{\mu}=\frac{1}{2}\bar{\Psi}\left(\nabla_{\mu}\beta^{\mu}\right)\Psi
\end{equation}
\noindent where $J^{\mu}=\frac{1}{2}\bar{\Psi}\beta^{\mu}\Psi$. The factor $1/2$ multiplying $\bar{\Psi}\beta^{\mu}\Psi$, of no importance regarding the conservation law, is in order to hand over a charge density conformable to that one used in the KG theory and its
nonrelativistic limit \cite{JPA43:055306:2010}. The adjoint spinor $\bar{\Psi}$ is given by $\bar{\Psi}=\Psi^{\dagger }\eta ^{0}$ with $\eta ^{0}=2\beta ^{0}\beta ^{0}-1$ in such a way
that $(\eta^{0}\beta^{\mu})^{\dag}=\eta^{0}\beta^{\mu}$ (the matrices $\beta^{\mu}$ are Hermitian with respect to $\eta^{0}$). Thus, if
\begin{equation}\label{cj0}
    \nabla_{\mu}\beta^{\mu}=0
\end{equation}
\noindent then four-current will be conserved. The condition (\ref{cj0}) is the purely geometrical assertion that the curved-space beta matrices are covariantly constant.

On the other hand, the normalization condition $\int{d\tau J^{0}}=\pm 1$ can be expressed as
\begin{equation}\label{normali1}
\int{d\tau \bar{\Psi}\beta^{0}\Psi}=\pm 2\,,
\end{equation}
\noindent where the plus (minus) sign must be used for a positive (negative) charge, and the expectation value of any observable $\mathcal{O}$ can be given by
\begin{equation}\label{veo}
\langle \mathcal{O} \rangle=\frac{\int d\tau \bar{\Psi}\beta^{0}\mathcal{O}\Psi}{\int d\tau \bar{\Psi}\beta^{0}\Psi}\,,
\end{equation}
\noindent where $\beta^{0}\mathcal{O}$ should be Hermitian with respect to $\eta^{0}$, $\left[\eta^{0}\left(\beta^{0}\mathcal{O}\right)\right]^{\dagger}=\eta^{0}\left(\beta^{0}\mathcal{O}\right)$, in order to provide real eigenvalues \cite{PRA90:022101:2014}.

\subsection{Interaction in the Duffin-Kemmer-Petiau equation}
\label{subsec:1}

With the introduction of interactions, the DKP equation in a curved space-time can be written as%
\begin{equation}
\left( i\beta ^{\mu }\nabla _{\mu }-M-U\right) \Psi =0  \label{dkp2}
\end{equation}
\noindent where the more general potential matrix $U$ is written in terms of
25 (100) linearly independent matrices pertinent to five (ten)-dimensional
irreducible representation associated to the scalar (vector) sector. The potential matrix $U$ can be written in terms of well-defined Lorentz structures. For the scalar sector (spin-zero) there are
two scalar, two vector and two tensor terms \cite{PRD15:1518:1977}, whereas for the vector sector (spin-one) there are two scalar, two vector, a pseudoscalar, two
pseudovector and eight tensor terms \cite{JPA12:665:1979}.

In the presence of interaction, $J^{\mu }$ satisfies the equation
\begin{equation}
\nabla _{\mu }J^{\mu }+\frac{i}{2}\,\bar{\Psi}\left( U-\eta ^{0}U^{\dagger
}\eta ^{0}\right) \Psi =\frac{1}{2}\bar{\Psi}\left(\nabla_{\mu}\beta^{\mu}\right)\Psi  \label{corrent2}
\end{equation}
\noindent Thus, if $U$ is Hermitian with respect to $\eta ^{0}$ and the curved-space beta matrices are covariantly constant then four-current will be conserved. The condition (\ref{corrent2}) for the case of Minkowski space-time has been used to point out a
misleading treatment in the recent literature regarding analytical solutions
for nonminimal vector interactions \cite{ADHEP2014:784072:2014}.

\section{Noninertial reference frame and the cosmic string background}
\label{sec:2}

The cosmic string space-time is an object described by the line element
\begin{equation}\label{met}
ds^{2}=- dT^{2} +dR^{2}+\alpha^{2}R^{2}d\Phi^{2}+dZ^{2}
\end{equation}
\noindent where $-\infty<Z<+\infty$, $R\geq 0$ and $0\leq\Phi\leq2\pi$. The parameter $\alpha$ is associated with the linear mass density $\tilde{m}$ of the string by $\alpha=1-4\tilde{m}$ and runs in the interval $\left( 0,1 \right]$ and corresponds to a deficit angle $\gamma=2\pi(1-\alpha)$. In the geometric context, the line element (\ref{metric}) is related to a Minkowski space-time with a conical singularity \cite{JHEP2004:016:2004}. Note that, in the limit as $\alpha\rightarrow1$ we obtain the line element of cylindrical coordinates.

The rotating frame is obtained using the following coordinate transformation,
\begin{equation}\label{rf}
T=t\,, \quad R=r\,,\quad \Phi=\varphi+\varpi t\,,\quad Z=z\,,
\end{equation}
\noindent where $\varpi$ is the constant angular velocity of the rotating frame. So, the line element (\ref{met}) becomes
\begin{equation}\label{metric}
\begin{split}
ds^{2}= & -\left( 1-\varpi^{2}\alpha^{2}r^{2} \right) dt^{2} + 2\varpi\alpha^{2}r^{2}d\varphi dt \\
&   +dr^{2}+\alpha^{2}r^{2}d\varphi^{2}+dz^{2}\,.
\end{split}
\end{equation}
\noindent This line element describes the background of a cosmic string in a rotating coordinate system. It is worthwhile to mention that the line element (\ref{metric}) is defined in the interval $0<r<r_{0}$, where $r_{0}=1/\varpi\alpha$ and that for values of $r>r_{0}$ correspond to a particle placed outside of the line cone. This interesting fact imposes one restriction on the radial coordinate, the wave function of the quantum particle must vanish at $r\rightarrow r_{0}$. This peculiar behavior can be interpreted of way that the geometry of the space-time plays the role of a hard--wall confining potential \cite{PRD82:084025:2010,PLA374:4642:2010,IJMPA26:4239:2011,MPLB27:1350018:2013,PRD89:027702:2014}.

The basis tetrad $e^{\mu}\,_{\bar{a}}$ from the line element (\ref{metric}) is chosen to be
\begin{equation}\label{tetra}
e^{\mu}\,_{\bar{a}}=%
\begin{pmatrix}
\frac{1}{\sqrt{1-\rho^{2}}} & 0 & \frac{\varpi\alpha r}{\sqrt{1-\rho^{2}}} &  0\\
0 & 1 & 0 & 0\\
0 & 0 & \frac{\sqrt{1-\rho^{2}}}{\alpha r} & 0\\
0 & 0 & 0 & 1
\end{pmatrix}%
\,,
\end{equation}%
\noindent where $\rho=\varpi\alpha r$. For the specific basis tetrad (\ref{tetra}) the curved-space beta matrices read
\begin{eqnarray}
\beta^{0} &=& \frac{1}{\sqrt{1-\rho^{2}}}\left( \beta^{\bar{0}}+\varpi\alpha r\beta^{\bar{2}} \right)\,,  \label{betat} \\
\beta^{r} &=& \beta^{\bar{1}}\,,   \label{betar} \\
\beta^{\varphi} &=& \frac{\sqrt{1-\rho^{2}}}{\alpha r}\beta^{\bar{2}}\,,  \label{betaphi}\\
\beta^{z}&=& \beta^{\bar{3}}\,, \label{betaz}
\end{eqnarray}%
\noindent and thereby, the covariant derivative gets
\begin{eqnarray}
\nabla_{0} &=& \partial_{0}-\Gamma_{0} \,,\label{dt}\\
\nabla_{r} &=& \partial_{r}-\Gamma_{r}  \,,\label{dr}\\
\nabla_{\varphi} &=& \partial_{\varphi}-\Gamma_{\varphi} \,,\label{dphi}\\
\nabla_{z} &=& \partial_{z}
\end{eqnarray}
\noindent where the spin connections are given by
\begin{equation}\label{g0}
\Gamma_{0}=\frac{\varpi\alpha}{\sqrt{1-\rho^{2}}}\left(\varpi\alpha r \left[\beta^{\bar{0}},\beta^{\bar{1}}\right]-\left[\beta^{\bar{1}},\beta^{\bar{2}}\right] \right) \,,
\end{equation}
\begin{equation}\label{gr}
\Gamma_{r}=-\frac{\varpi\alpha}{1-\rho^{2}} \left[\beta^{\bar{0}},\beta^{\bar{2}}\right] \,,
\end{equation}
\begin{equation}\label{gphi}
\Gamma_{\varphi}=\frac{\Gamma_{0}}{\varpi} \,.
\end{equation}
\noindent Note that using the line element (\ref{metric}) and the representation for the curved-space beta matrices (\ref{betat}), (\ref{betar}), (\ref{betaphi}) and (\ref{betaz}) the condition (\ref{cj0}) is satisfied and therefore the current is conserved for this background.

\section{DKP oscillator in a noninertial reference frame}
\label{sec:3}

In this section, we concentrate our efforts in the interaction called DKP oscillator. For this external interaction we use the non-minimal substitution \cite{JPA27:4301:1994}
\begin{equation}
\vec{p}\rightarrow \vec{p}-iM\omega\eta^0\vec{r}
\end{equation}
\noindent where $\omega$ is the oscillator frequency. This interaction is a Lorentz-tensor type and is Hermitian with respect to $\eta^0$, so it furnishes a conserved four-current. Considering only the radial component the non-minimal substitution gets
\begin{equation}
\vec{p}\rightarrow \vec{p}-iM\omega\eta^0r\hat{r}\,.
\end{equation}
\noindent As the interaction is time-independent one can write $\Psi(\vec{r},t)=\Phi(\vec{r})\mathrm{exp}\left(-iEt\right)$, where $E$ is the energy of the scalar boson, in such a way that the time-independent DKP equation becomes
\begin{equation}
\begin{split}
\left[ \beta^{0}\left(E-\Gamma_{0}\right)+
 i\beta^{\bar{1}}\left( \partial_{r}-\Gamma_{r}+M\omega\eta^{0}r \right) \right.\\
 \left.+ i\beta^{\varphi}\left(\partial_{\varphi}-\Gamma_{\varphi}\right)+
i\beta^{\bar{3}}\partial_{z}-M \right]\Phi =0
\end{split}
\end{equation}
\noindent where $\beta^{0}$, $\beta^{\varphi}$, $\Gamma_{0}$, $\Gamma_{r}$ and $\Gamma_{\varphi}$ are given by (\ref{betat}), (\ref{betaphi}), (\ref{g0}), (\ref{gr}) and (\ref{gphi}), respectively.

\subsection{Scalar sector}
\label{subsec:3}

For the case of scalar bosons (scalar sector), we use the standard representation for the beta matrices given by \cite{JPG19:87:1993}
\begin{equation}
\beta ^{\bar{0}}=%
\begin{pmatrix}
\theta & \overline{0} \\
\overline{0}^{T} & \mathbf{0}%
\end{pmatrix}%
,\quad \overrightarrow{\beta }=%
\begin{pmatrix}
\widetilde{0} & \overrightarrow{\sigma } \\
-\overrightarrow{\sigma }^{\,T} & \mathbf{0}%
\end{pmatrix}%
\end{equation}%
\noindent where%
\begin{eqnarray}
\ \theta &=&%
\begin{pmatrix}
0 & 1 \\
1 & 0%
\end{pmatrix}%
,\quad \sigma ^{1}=%
\begin{pmatrix}
-1 & 0 & 0 \\
0 & 0 & 0%
\end{pmatrix}
\notag \\
&& \\
\sigma ^{2} &=&%
\begin{pmatrix}
0 & -1 & 0 \\
0 & 0 & 0%
\end{pmatrix}%
,\quad \sigma ^{3}=%
\begin{pmatrix}
0 & 0 & -1 \\
0 & 0 & 0%
\end{pmatrix}
\notag
\end{eqnarray}
\noindent $\overline{0}$, $\widetilde{0}$ and $\mathbf{0}$ are 2$\times $3, 2%
$\times $2 and 3$\times $3 zero matrices, respectively, while the
superscript T designates matrix transposition. The five-component spinor can be written as $\Phi ^{T}=\left( \Phi_{1},...,\Phi _{5}\right)$ and the DKP equation for scalar bosons becomes
\begin{equation}\label{phi1}
\begin{split}
\frac{E}{\sqrt{1-\rho^{2}}}\Phi_{2}-M\Phi_{1}-i\left(\partial_{-}+\frac{1}{r}\right)\Phi_{3}-i\partial_{z}\Phi_{5} \\
-\frac{\varpi\alpha r}{M\sqrt{1-\rho^{2}}} \left( E+\frac{i\left(1-\rho^{2}\right)}{\varpi\alpha^{2}r^{2}}\partial_{\varphi} \right)\Phi_{4}=0\,,
\end{split}
\end{equation}
\begin{equation}\label{phi2}
\Phi_{2} = \frac{E}{M\sqrt{1-\rho^{2}}}\Phi_{1} \,,
\end{equation}
\begin{equation}\label{phi3}
\Phi_{3} = \frac{i}{M} \partial_{+}\Phi_{1} \,,
\end{equation}
\begin{equation}\label{phi4}
\Phi_{4} = \frac{\varpi\alpha r}{M\sqrt{1-\rho^{2}}} \left( E+\frac{i\left(1-\rho^{2}\right)}{\varpi\alpha^{2}r^{2}}\partial_{\varphi} \right)\Phi_{1}  \,,
\end{equation}
\begin{equation}\label{phi5}
\Phi_{5} = \frac{i}{M}\partial_{z}\Phi_{1} \,,
\end{equation}
\noindent where
\begin{eqnarray}
\partial_{-}&=&\partial_{r}-M\omega r \,,\label{dplus}\\
\partial_{+}&=&\partial_{r}+M\omega r \,, \label{dminus}
\end{eqnarray}
\noindent Meanwhile,
\begin{equation}\label{norma2}
\begin{split}
J^{0}=\frac{1}{\sqrt{1-\rho^{2}}}\left[\mathrm{Re}\left(\Phi_{2}^{\ast}\Phi_{1}\right)-
\rho\mathrm{Re}\left(\Phi_{4}^{\ast}\Phi_{1}\right)\right]\\
=\frac{E|\Phi_{1}|^{2}+\varpi\mathrm{Re}\left(i\Phi_{1}\partial_{\varphi}\Phi_{1}^{\ast}\right)}{M}\,.
\end{split}
\end{equation}
\noindent Combining these results we obtain a equation of motion for the first component of the DKP spinor
\begin{equation}\label{ed2o1}
\begin{split}
\left[\nabla_{\alpha}^2 \right. & \left. -M^2\omega^2r^2-2iE\varpi\partial_{\varphi}-\varpi^{2}\partial_{\varphi}^{2}\right. \\
& \left.+E^2-M^2+2M\omega\right]\Phi_{1}=0
\end{split}
\end{equation}
\noindent where $\nabla_{\alpha}^2$ is the Laplacian-Beltrami operator in the conical space, given by
\begin{equation}\label{laplacc}
\nabla_{\alpha}^2=\frac{1}{r}\frac{\partial}{\partial r}\left(r\frac{\partial}{\partial r}\right)+
\frac{1}{\alpha^2r^2}\frac{\partial^2}{\partial \varphi^2}+\frac{\partial^2}{\partial z^2}\,.
\end{equation}
\noindent At this stage, we can use the invariance under boosts along the $z$-direction and adopt the usual decomposition
\begin{equation}\label{ansatz}
\Phi_{1}(r,\varphi,z)=\frac{\phi_{1}(r)}{\sqrt{r}}e^{il\varphi+ik_{z}z}
\end{equation}
\noindent with $l=0,\pm1,\pm2,\ldots$\,. Inserting this into Eq. (\ref{ed2o1}), we get
\begin{equation}\label{ed2o}
\left[\frac{d^2}{dr^2}-\lambda^2r^2-\frac{\left(l_{\alpha}^2-\frac{1}{4}\right)}{r^2}+\kappa^2\right]\phi_{1}=0
\end{equation}
\noindent where $l_{\alpha}=l/\alpha$, $\lambda=M\omega$ and
\begin{equation}\label{kappa}
\kappa=\sqrt{\left(E+\varpi l\right)^2-M^2+2M\omega-k_{z}^{2}}\,.
\end{equation}
\noindent The motion equation (\ref{ed2o}) describes the quantum dynamics of a DKP oscillator in the backgroud of a cosmic string in a rotating coordinate system. The solution close to the origin valid for all values de $l_{\alpha}$ can be written as being proportional to $r^{|l_{\alpha}|+\frac{1}{2}}$. On the other hand, for sufficiently large radius $r_{0}$ the square-integrable solution behaves as
$e^{-\lambda r^2/2}$, thereby the solution for $0<r<r_{0}$ can be expressed as
\begin{equation}\label{ansatz2}
\phi_{1}(r)=r^{|l_{\alpha}|+\frac{1}{2}}e^{-\lambda r^2/2}f(r)\,,
\end{equation}
\noindent subsequently, by introducing the following new variable and parameters:
\begin{eqnarray}
 \xi &=& \lambda r^2\,, \label{newvar}\\
 a &=& \frac{1}{2}\left(|l_{\alpha}|+1-\frac{\kappa^2}{2\lambda}\right)\,,\label{a}\\
 b &=& |l_{\alpha}|+1\,,\label{b}
\end{eqnarray}
\noindent one finds that $f(\xi)$ can be expressed as a regular solution of the confluent hypergeometric equation (Kummer's function) \cite{ABRAMOWITZ1965},
\begin{equation}\label{hipergeome}
\xi\frac{d^2f}{d\xi^2}+\left(b-\xi\right)\frac{df}{d\xi}-af=0\,.
\end{equation}
\noindent The general solution of (\ref{hipergeome}) is given by \cite{ABRAMOWITZ1965}
\begin{equation}\label{gs}
f(\xi)=AM\left(a,b,\xi\right)+B\xi^{1-b}M\left(a-b+1,2-b,\xi\right)
\end{equation}
\noindent where $A$ and $B$ are arbitrary constants. The second term in (\ref{gs}) has a singular point at $\xi=0$, so that we set $B=0$. In such a way that the solution for (\ref{hipergeome}) is given by
\begin{equation}\label{solhi}
f(\xi)=AM\left(a,b,\xi\right)
\end{equation}
\noindent As mentioned in the Sect. \ref{sec:2}, the peculiar behavior of this background which is defined in the interval $0<r<r_{0}$, where $r_{0}=1/\varpi\alpha$, the problem presents two different classes of solutions that depend on the value of the product $\varpi\alpha$. Let's us consider as a first case an arbitrary value of $\varpi\alpha$ and as a second case the limit $\varpi\alpha\ll 1$. At the two next sections, we will analyze each case in detail.

\subsection{ Arbitrary $\varpi\alpha$}
\label{subsec:3:2}

Following the discussions of the Sect. \ref{sec:2}, we proceed now to find the eigenfunction for this problem. Because the restriction on the radial coordinate due to noninertial effects a physical solution is possible only if the eigenfunction vanishes at $r=r_{0}=1/\varpi\alpha$ in order to normalize $\phi_{1}$, thereby the boundary condition implies that
\begin{equation}\label{bc}
M\left(a,b,\frac{\lambda}{\varpi^{2}\alpha^{2}}  \right)=0\,.
\end{equation}
\noindent By solving this quantization condition one obtains the possible energy levels by inserting the allowed values of $a=a_{l}$ in (\ref{a}) and combining with (\ref{kappa}) yields
\begin{equation}\label{energia1}
E_{\pm}=\pm\sqrt{2M\omega\left( \frac{|l|}{\alpha}-2a_{l} \right)+M^{2}+k_{z}^{2}}-\varpi | l |\,,
\end{equation}
\noindent which is irrespective to the sign of angular momentum quantum number $|l|$. From (\ref{energia1}) we can see that the discrete set of DKP energies is composed of two contributions: the first term of (\ref{energia1}) is associated to the DKP oscillator embedded in a cosmic string background and the second term of (\ref{energia1}) is associated to the noninertial effect of rotating frames, which in turn is a Sagnac-type effect \cite{CRAS157:708:1913,CRAS157:1410:1913}. Note that both particle ($E_{+}$) and antiparticle ($E_{-}$) energy levels are members of the spectrum and also that the noninertial effect is to break the symmetry of the energy spectrum about $E=0$. From (\ref{energia1}) we can conclude that $|E_{-}|>|E_{+}|$. Furthermore, if $|l|=0$ or $\varpi=0$ the discrete set of DKP energies are symmetrical about $E=0$. At this stage, we can use the invariance under boosts along the $z$-direction and without loss of generality we can fix $k_{z}=0$.

Although the quantization condition has no closed form solutions in terms of simpler functions, the numerical computation of $a_{l}$ can be done easily with a root-finding procedure of a symbolic algebra program. The first values of $a_{l}$ that satisfy the quantization condition (\ref{bc}) and its respective energies are listed in Tables \ref{table1} and \ref{table2} for $\varpi=0.5$ and $\varpi=1.0$, respectively. 

\begin{figure}[h]
\includegraphics[width=0.93\linewidth,angle=0]{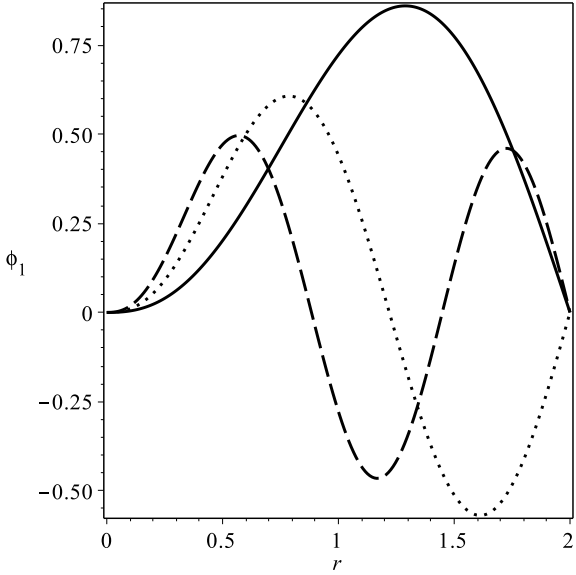}
\caption{\label{fig1} (Color online) Plots of $\phi_{1}$ (normalized) for the ground state (solid line),  first excited state (dotted line) and second excited state (dashed line) for $\omega=0.1$, $|l|=1$, $\alpha=0.5$, and $\varpi=1.0$.}
\end{figure}

With all that, the solution for $0<r<r_{0}$ can be written as
\begin{equation}\label{solution}
\phi_{1}\left(r\right)=A_{l}\,r^{|l_{\alpha}|+\frac{1}{2}}e^{-\lambda r^2/2}M(a_{l},b,\lambda r^{2})\,,
\end{equation}
\noindent where $A_{l}$ is a normalization constant. The charge density $J^{0}$ (\ref{norma2}) dictates that
$\phi_{1}$ must be normalized as
\begin{equation}\label{normali3}
\frac{|E_{\pm}+\varpi |l||}{M}\int_{0}^{r_{0}}dr|\phi_{1}|^{2}=1\,,
\end{equation}
\noindent so that, the normalization constant can be written as
\begin{equation}\label{normali4}
A_{l}=\sqrt{\frac{M}{\delta |E_{\pm}+\varpi |l||}}\,,
\end{equation}
\noindent where
\begin{equation}\label{delta}
\delta=\int_{0}^{r_{0}}dr\, r^{2|l_{\alpha}|+1}e^{-\lambda r^{2}}
|M\left(a_{l},b,\lambda r^{2}\right)|^{2}
\end{equation}
\noindent with $|E_{\pm}+\varpi |l||\neq 0$. From (\ref{normali4}) we can see that, as $|E_{+}+\varpi |l||=|E_{-}+\varpi |l||$ is expected that $A_{l}$ for particles is equal to $A_{l}$ for antiparticles. Figure \ref{fig1} illustrates the behavior of $\phi_{1}$ (normalized) for the three lowest states, $\omega=0.1$, $|l|=1$, $\alpha=0.5$ and $\varpi=1.0$. Here, we consider only bosons, i.e. $E_{+}$. Note that $\varpi\alpha=0.5$, so we have that the solution is restricted to the interval $0<r<2$. Also, we can note that the solution for the ground state has no nodes, the first excited state has one node and the second excited state has two nodes. From this fact, we can conclude that exist a relation systematic between number of nodes of $\phi_{1}$ and each level of states (usual node structure), even if there is a restriction at the radial variable. In other context, a similar restriction in an one-dimensional problem was studied in \cite{AP351:571:2014}.

\begin{figure}[h]
\includegraphics[width=0.93\linewidth,angle=0]{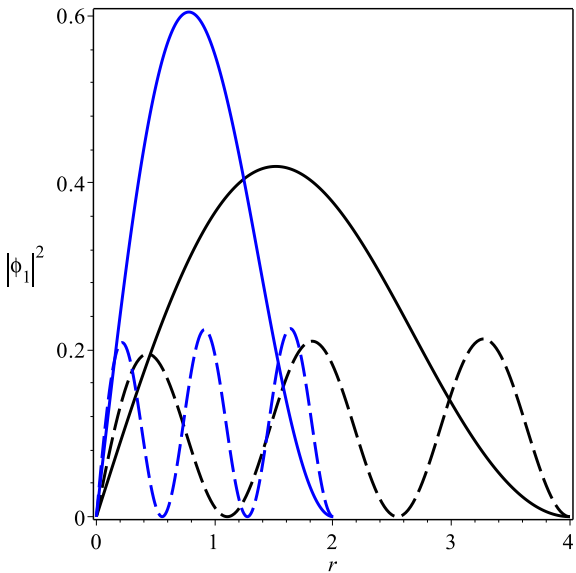}
\caption{\label{fig2} (Color online) Plots of $|\phi_{1}|^{2}$ for the ground state (solid line) and the second excited state (dashed line), $\omega=0.1$, $|l|=0$, $\alpha=0.5$, and for
$\varpi=0.5$ (black) and $\varpi=1.0$ (blue).}
\end{figure}

\begin{figure}[h]
\includegraphics[width=0.93\linewidth,angle=0]{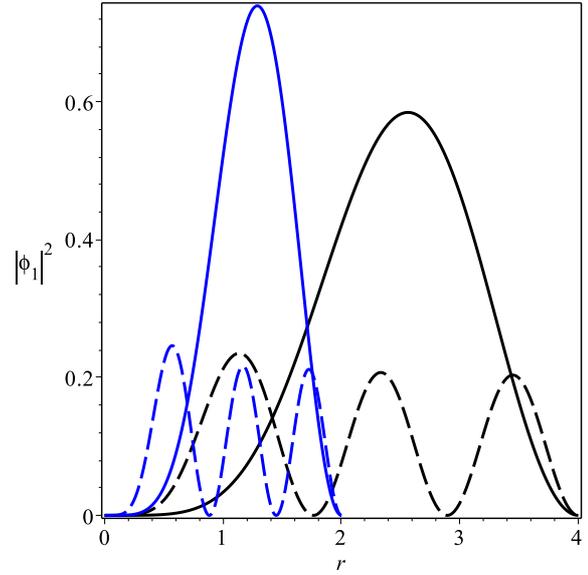}
\caption{\label{fig3} (Color online) The same as Fig. \ref{fig2}, for $|l|=1$.}
\end{figure}

In Figs. \ref{fig2} and \ref{fig3}, we illustrate the results of $|\phi_{1}|^{2}$ for the ground state and the second excited state, $\omega=0.1$, $\alpha=0.5$ and two different values of $\varpi$. Figures \ref{fig2} and \ref{fig3} clearly show the noninertial effects on the excited states, which are qualitatively similar to $|l|=0$ and $|l|=1$, respectively. From Fig. \ref{fig2} one can see that for the ground state, the distribution has a maximum at $r\approx 1.5$ for $\varpi=0.5$ and $|l|=0$, and this maximum increases and moves to negative r-direction as $\varpi=1.0$. 

Then, for study the scalar bosons localization we use (\ref{veo}), the quantity $ \left(\Delta x\right)^{2}=\langle x^{2}\rangle-\langle x\rangle^{2} $  can be written as \cite{JPA43:055306:2010}
\begin{equation}\label{deltax}
\left(\Delta x\right)^{2}=\int_{0}^{r_{0}}dr|J^{0}|^{2}r^{2}-\left(\int_{0}^{r_{0}}dr|J^{0}|r\right)^{2}\,.
\end{equation}
\noindent Calculating $\Delta x$ for the parameter set of figure \ref{fig2} ($\omega=0.1$, $|l|=0$ and $\alpha=0.5$), we obtain $\Delta x=0.7773$ ($\varpi=0.5$) and $\Delta x=0.3909$ ($\varpi=1.0$) for the ground state and $\Delta x=1.1477$ ($\varpi=0.5$) and $\Delta x=0.5740$ ($\varpi=1.0$) for the second excited state. A similar behavior is observed at $\Delta x$ for the parameter set of figure \ref{fig3} ($\omega=0.1$, $|l|=1$ and $\alpha=0.5$). In this case, we obtain $\Delta x=0.5904$ ($\varpi=0.5$) and $\Delta x=0.2946$ ($\varpi=1.0$) for the ground state and $\Delta x=1.0019$ ($\varpi=0.5$) and $\Delta x=0.5005$ ($\varpi=1.0$) for the second excited state. From these results, we can conclude that scalar bosons tend to be better localized when $\varpi$ increases. If instead of considering bosons, we consider antibosons (i.e. $E_{-}$), one would expect the same results, since $A_{l}$ for particles is equal to $A_{l}$ for antiparticles.

\begin{table}[ht]
\centering
\begin{tabular}{c|cc|ccc}
\hline
\multicolumn{6}{ c }{$M\left(a=a_{l},b,\frac{\lambda}{\varpi^{2}\alpha^{2}}\right)=0$}\\
\hline
               & \multicolumn{2}{ |c| }{$l=0$} & \multicolumn{3}{ |c }{$l=1$} \\
\cline{2-6}
 $\alpha$ &  $a_{l}$ &  $|E_{\pm}|$ &  $a_{l}$  & $E_{+}$ & $E_{-}$ \\
\hline
  $0.9$ & -2.4546 & 1.4078  & -7.0092 & 1.5065 & -2.5065  \\
                  & -14.9645 & 2.6431 & -25.0484 & 2.8528 & -3.8528 \\
                   & -37.4516 & 3.9976  & -53.0832 & 4.2387 & -5.2387 \\
                  & -69.9297 & 5.3826 & -91.1117 & 5.6373 & -6.6373 \\
               & -112.4003 & 6.7794 & -139.1333 & 7.0416  & -8.0416 \\
\hline
   $0.5$ &  -0.4896  & 1.0935   & -2.7840 & 1.0854 & -2.0854 \\
                        & -4.3861 & 1.6596 & -9.7150 &  1.7991  & -2.7991 \\
                      &  -11.3311 & 2.3521  & -19.7364 & 2.5487  & -3.5487 \\
                      & -21.3566 & 3.0891 & -32.8433 & 3.3128  & -4.3128 \\
                      & -34.4654 & 3.8453 & -49.0351 & 4.0841  & -5.0841 \\
\hline
  $0.1$ & 0.0000 &  1.0000 & 0.0000 & 1.2321  & -2.2321 \\
                      & -1.0000  & 1.1832 & -1.0000 & 1.3439 & -2.3439 \\
                     & -2.0000 & 1.3416 & -2.0006 & 1.4494 & -2.4494 \\
                     & -3.0000 & 1.4832 & -3.0075 & 1.5501  & -2.5501 \\
                    & -4.0000 & 1.6125 & -4.0476 & 1.6492 & -2.6492 \\
\hline
\end{tabular}
\caption{The first values of $a_{l}$ that satisfy the quantization condition (\ref{bc}) for $\omega=0.1$ and $\varpi=0.5$.}
\label{table1}
\end{table}

\begin{table}[ht]
\centering
\begin{tabular}{c|cc|ccc}
\hline
\multicolumn{6}{ c }{$M\left(a=a_{l},b,\frac{\lambda}{\varpi^{2}\alpha^{2}}\right)=0$}\\
\hline
      & \multicolumn{2}{ |c| }{$l=0$} & \multicolumn{3}{ |c }{$l=1$} \\
\cline{2-6}
   $\alpha$   &  $a_{l}$ &  $|E_{\pm}|$  &  $a_{l}$  & $E_{+}$ & $E_{-}$ \\
\hline
  $0.9$ & -11.2177 & 2.3425 & -31.0448  & 2.6933 & -4.6933  \\
             & -61.2139 &  5.0483   & -103.2043 & 5.5195 & -7.5195  \\
             & -151.1562 & 7.8398   & -215.3445 & 8.3467 & -10.3467 \\
             & -281.0667 & 10.6502  & -367.4586 & 11.1740 & -13.1740 \\
             & -450.9481 & 13.4677 & -559.5451 & 14.0013 & -16.0013  \\
\hline
   $0.5$    &  -3.1363  & 1.5015   & -15.0250 & 1.7221 & -3.7221 \\
               & -18.5757 & 2.9035 & -42.8174 &  3.3043 & -5.3043  \\
              &  -46.3368 & 4.4198  & -82.9228 & 4.8795 & -6.8795  \\
              & -86.4330 & 5.9643 & -135.3594 & 6.4528  & -8.4528 \\
               & -138.8657 & 7.5197 & -200.1309 & 8.0251 & -10.0251 \\
\hline
  $0.1$    & -0.0004 &  1.0001  & -1.3258 & 0.8789 & -2.8789  \\
            &  -1.0238  & 1.1872  & -4.3039 & 1.1729  & -3.1729 \\
             & -2.2262 & 1.3750  & -7.8222 & 1.4757  & -3.4757 \\
            & -3.8315 & 1.5914   & -11.8557 & 1.7825  & -3.7825 \\
              & -5.9225 & 1.8355  & -16.3947 & 2.0916 & -4.0916 \\
\hline
\end{tabular}
\caption{The first values of $a_{l}$ that satisfy the quantization condition (\ref{bc}) for $\omega=0.1$, and $\varpi=1.0$.}
\label{table2}
\end{table}


\subsection{Limit $\varpi\alpha\ll 1$ ($r_{0}\rightarrow\infty$)}
\label{subsec:3:3}

The second class of solutions is obtained adopting the limit $\varpi\alpha\ll 1$. The main consequence is change to the boundary condition and the restriction on the radial coordinate due to noninertial effects. Considering a large $r_{0}$ ($r_{0}\rightarrow\infty$) the square-integrable solution is guaranteed by the term $e^{-\lambda r^2/2}$ in (\ref{ansatz2}) and without loss generality the boundary condition becomes
\begin{equation}\label{bc3}
M\left(a,b,\frac{\lambda}{\varpi^{2}\alpha^{2}}\rightarrow\infty\right)=\mathrm{finite}\,.
\end{equation}
\noindent Now we need to analyze the asymptotic behavior of the solution (\ref{solhi}).

The asymptotic behavior of Kummer's function is dictated by
\begin{equation}\label{asymp}
M\left(a,b,\xi\right)\simeq \frac{\Gamma(b)}{\Gamma(b-a)}e^{-i\pi a}\xi^{-a}+
\frac{\Gamma(b)}{\Gamma(a)}e^{\xi}\xi^{a-b}\,.
\end{equation}
\noindent It is true that the presence of $e^{\xi}$ in the second term of (\ref{asymp}) perverts the normalizability of $\phi_{1}(\xi)$ in (\ref{ansatz2}). Nevertheless, this unfavorable behavior can be remedied by demanding $a=-n$, where $n$ is a nonnegative integer and $b\neq-\tilde{n}$, where $\tilde{n}$ is also a nonnegative integer. As a matter of fact, $M(-n,b,\xi)$ with $b>0$ is proportional to the generalized Laguerre polynomial $L_{n}^{(b-1)}(\xi)$, a polynomial of degree $n$ with $n$ distinct positive zeros in the range $[0,\infty)$. Therefore, the solution for all $r$ can be written as
\begin{equation}\label{solg}
\phi_{1}(r)=A_{n}r^{|l_{\alpha}|+\frac{1}{2}}e^{-\lambda r^2/2}L^{|l_{\alpha}|}_{n}(\lambda r^{2})\,,
\end{equation}
\noindent where $A_{n}$ is a normalization constant. One more time, the charge density $J^{0}$ (\ref{norma2}) dictates that
$\phi_{1}$ must be normalized as
\begin{equation}\label{norma3}
\frac{|E_{\pm}+\varpi |l||}{M}\int_{0}^{\infty}dr|\phi_{1}|^{2}=1\,,
\end{equation}
\noindent so that, the normalization constant can be written as
\begin{equation}\label{norma4}
A_{n}=\sqrt{\frac{2M\lambda^{|l_{\alpha}|+1}\Gamma\left(n+1\right)}{|E_{\pm}+\varpi |l||\Gamma\left(|l_{\alpha}|+n+1\right)}}\,,
\end{equation}
\noindent with $|E_{\pm}+\varpi |l||\neq 0$. Moreover, the requirement $a=-n$ (quantization condition) implies into
\begin{equation}\label{energyr}
E_{\pm}=\pm\sqrt{2M\omega\left(2n+\frac{|l|}{\alpha}\right)+M^2+k_{z}^{2}}-\varpi |l|\,.
\end{equation}
\noindent Similar to the case of arbitrary $\varpi\alpha$, the discrete set of DKP energies is modified by the term $\varpi |l|$. This last result shows that the discrete set of DKP energies are not symmetrical about $E=0$. Here are applicable the same discussion already done for the energy spectrum (\ref{energia1}). As a particular case, making $\varpi=0$ we obtain the DKP energies of the DKP oscillator embedded in the background of a cosmic string, already reported in \cite{EPJC75:287:2015}. One more time, due to invariance under rotation along the z-direction, without loss of generality we can fix $k_{z}=0$.

Now, let us consider the nonrelativistic limit of (\ref{energyr}). Following the standard procedure, $E=M+\mathcal{E}$ with $M\gg\mathcal{E}$, and after some calculations one has that
\begin{equation}\label{lnr}
\mathcal{E}\simeq \omega\left(2n+\frac{|l|}{\alpha}\right)-\varpi |l|\,.
\end{equation}
\noindent This last result describes the energy of a traditional nonrelativistic harmonic oscillator plus the Page--Werner term \cite{PRL35:543:1975,PRL42:1103:1979}.

\begin{figure}[h]
\includegraphics[width=0.93\linewidth,angle=0]{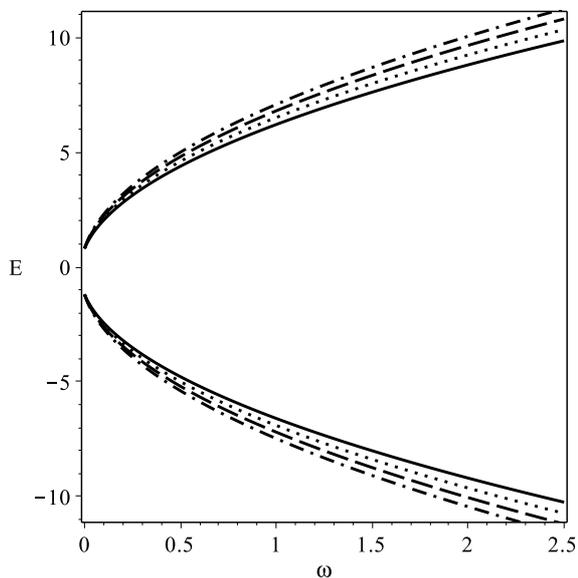}
\caption{\label{figE} (Color online) Plots of the energy as a function of $\omega$ for $|l|=2$, $\varpi=0.1$, $\alpha=0.1$, and different values of $n$. For $n=0$ (solid line), $n=1$ (dotted line), $n=2$ (dashed line), and $n=3$ (dot-dashed line).}
\end{figure}

Figure \ref{figE} illustrates the profile of the energy as a function of $\omega$ for $|l|=2$. In this figure we consider the four first quantum numbers. From figure \ref{figE} one sees that all the energy levels emerge from the positive (negative)-energy continuum so that it is plausible to identify them with particle (antiparticle) levels. 

Finally, our results for the limit $\varpi\alpha\ll 1$ are consistent with our results for arbitrary $\varpi\alpha$, this fact can be seen from Table \ref{table1}. In this table, the product $\varpi\alpha=0.05$ can be considered as $\ll 1$, because $a_{l}\rightarrow-n$ for $|l|=0$ or $|l|=1$, where $n$ is a nonnegative integer. This means that the choice of $\varpi=0.5$ and $\alpha=0.1$ that produce $r_{0}=20$ is suitable, because this $r_{0}$ can be considered as a sufficiently large radius.

\section{Conclusions}
\label{sec:4}

We studied the Duffin-Kemmer-Petiau (DKP) equation in a cosmic string background in a rotating reference frame. We showed that considering this background and a DKP oscillator interaction, they furnish a conserved current. Considering only scalar bosons, we calculated the motion equation, which describes the quantum dynamics of a DKP oscillator in this background and discussed in details the combined effects of the angular velocity of the rotating frame $\varpi$ and the angular deficit of the cosmic string background $\alpha$. This problem was mapped into a confluent hypergeometric equation for the first component of the DKP spinor $\phi_{1}$ and the remaining components were expressed in terms of the first one in a simple way. 

Due to peculiar behavior of this background, which is defined in the interval $0<r<r_{0}$, where $r_{0}=1/\varpi\alpha$, the problem presents two kinds of solutions that depend on the value of the product $\varpi\alpha$. As a first case, we considered an arbitrary value of $\varpi\alpha$. Using the appropriate boundary condition at $r=r_{0}$, we obtained the possible energy levels by a root-finding procedure of a symbolic algebra program. The first component of the DKP spinor $\phi_{1}$ was expressed in terms of the Kummer's function. As a second case, we considered the limit $\varpi\alpha\ll 1$, which implies $r_{0}\rightarrow\infty$ (a sufficiently large radius). One more time, using the appropriate boundary condition at $r=r_{0}\rightarrow\infty$, we obtained the energy levels in a analytic form. In this case, $\phi_{1}$ was expressed in terms of the generalized Laguerre polynomial. In both kinds of solutions, we showed that exist a relation systematic between number of nodes of $\phi_{1}$ and each level of states (usual node structure), even if there is a restriction at the radial variable.

For both kinds of solutions, we found the energy spectrum for this problem. We showed that the discrete set of DKP energies is composed of two contributions. One term is associated to the DKP oscillator embedded in a cosmic string background and the other term is associated to the noninertial effect of rotating reference frames, which in turn is a Sagnac-type effect. Both particle ($E_{+}$) and antiparticle ($E_{-}$) energy levels are members of the spectrum, and also that the noninertial effect breaks the symmetrical of the energy spectrum about $E=0$. Only for $ |l|=0 $ the energy spectrum is symmetrical about $E=0$. In general, we showed that $|E_{-}|>|E_{+}|$ and that bosons as well as antibosons tend to be better localized than $\varpi$ increases. We obtained the results reported in \cite{EPJC75:287:2015} as a particular case making $\varpi\rightarrow 0$. We also found that the nonrelativistic limit furnishes the energy of a traditional nonrelativistic harmonic oscillator plus the Page--Wenner term. Further, we showed that both kinds of solutions are consistent when a suitable $r_{0}$ (sufficiently large) is chosen.

Beyond to investigate the quantum dynamics of scalar bosons, the results of this paper could be used, in principle, in condensed matter physics. It is known the analogy between cosmic strings and disclinations in solids \cite{NELSON2002}, this fact is associated to that the metric which describes a disclination corresponds to the spatial part of the line element of the cosmic string. Thereby, our results could be used to investigate the integer quantum Hall effect for bosons \cite{PRL110:046801:2013} in a system with the presence of a topological defect in a rotating frame as done in \cite{EPL54:502:2001} for the quantum Hall effect, and could also be used to investigate symmetry-protected topological (SPT) phase \cite{PRL115:116802:2015}, which is the analogues of the celebrated free fermion topological insulators for bosonic systems. In this context, the DKP theory has been employed on the study of novel topological semimetals \cite{PRB92:235106:2015}. Another physical application could be associated to Bose-Einstein (BE) condensates \cite{PLA316:33:2003,PA419:612:2015} and neutral atoms, which can be used to study entanglement and quantum information processing \cite{PRL82:1975:1999}. Specifically with respect to BE condensates the idea is to rotate the BE condensate and to observe the generation of vortices as done in \cite{PRL76:6:1996,PRA55:2126:1997}.

\begin{acknowledgements}
The author would like to thank the referees for useful comments and suggestions. 
This work was supported in part by means of funds provided by CNPq, Brazil, Grants No. 455719/2014-4 (Universal) and No. 304105/2014-7 (PQ).
\end{acknowledgements}

\bibliographystyle{spphys}       
\bibliography{mybibfile_stars2}

\end{document}